\documentclass{article}

\PassOptionsToPackage{numbers, compress}{natbib}

\usepackage[preprint]{neurips_2026}

\usepackage[utf8]{inputenc} 
\usepackage[T1]{fontenc}    
\usepackage{hyperref}       
\usepackage{url}            
\usepackage{booktabs}       
\usepackage{amsfonts}       
\usepackage{amsmath}        
\usepackage{amssymb}        
\usepackage{nicefrac}       
\usepackage{microtype}      
\usepackage{xcolor}         
\usepackage{graphicx}       
\usepackage{subcaption}     

\graphicspath{{figures/}{../figures/}{../talk/assets/}}

\title{A Differentiable Neural Surrogate for Photon Propagation in Neutrino Telescopes}

\author{%
  Felix J.~Yu$^{1,2}$ \quad Berthy T.~Feng$^{3,2}$ \quad Nicholas Kamp$^{1,2}$ \quad Carlos A.~Arg\"{u}elles$^{1,2}$ \\[3pt]
  $^1$Harvard University \quad $^2$NSF IAIFI \quad $^3$Massachusetts Institute of Technology \\[2pt]
  \texttt{\{felixyu,nkamp,carguelles\}@g.harvard.edu}, \texttt{berthy@mit.edu} \\
}

\begin{document}

\maketitle

\begin{abstract}
Large-volume neutrino telescopes infer neutrino properties from Cherenkov light, but simulating the transport of billions of photons through highly scattering ice or water is computationally costly.
We introduce \texttt{candela}, a differentiable SIREN neural field that learns the photon Green's function of the IceCube Neutrino Observatory, a cubic-kilometer detector embedded in Antarctic glacial ice.
Given a point-like energy deposit and sensor, it predicts the expected photon yield and full arrival-time distribution at the sensor. 
Complete events are simulated by decomposing charged-particle energy deposits into point-like sources and superposing their predicted sensor responses.
Trained on Monte-Carlo simulations, \texttt{candela} generates events $50$--$100\times$ faster than existing methods, with cost scaling only weakly with neutrino energy.
It keeps median yields within $2\%$ of the MC expectation and timing distributions at the MC statistical floor across six photon-count decades.
The model also provides end-to-end gradients with respect to event parameters and opens a path toward optimizing scattering-medium properties, which often dominate systematic uncertainties in neutrino telescopes.
\end{abstract}

\section{Introduction}

The IceCube Neutrino Observatory is a cubic-kilometer neutrino telescope designed to detect high-energy neutrinos from astrophysical sources. 
When a neutrino interacts in or near the detector, the resulting charged particles produce Cherenkov light that is recorded by photosensors distributed throughout the instrumented volume. 
These particles include leptons such as muons and electrons, as well as hadrons.
Muons, heavier relatives of electrons, can travel long distances and produce elongated \textit{tracks}, whereas electrons and hadrons typically deposit their energy locally in showers called \textit{cascades}.
The amount and timing of the detected light are used to infer the properties of the original neutrino, making high-fidelity simulations of photon propagation central to IceCube's scientific program.
Such simulations underpin tasks such as particle track reconstruction, background event classification, and sensor design.
The simulation burden is amplified by IceCube's extreme event-rate imbalance: atmospheric muons trigger the detector at $2.5$--$2.9\,\mathrm{kHz}$, whereas selected atmospheric-neutrino and high-purity astrophysical-neutrino samples occur at roughly $\mathrm{mHz}$ and $\mu\mathrm{Hz}$ rates, respectively~\citep{icecube2017instrumentation}.
Accurately estimating the small fraction of atmospheric-muon events that survive neutrino selections therefore requires enormous simulated background samples.

IceCube achieves its enormous detection volume by instrumenting naturally occurring glacial ice between approximately 1.5 and 2.5 km beneath the surface of the South Pole~\citep{icecube2017instrumentation}. 
This introduces substantial complexity into its optical response: spatially varying dust concentrations, depth-dependent optical properties, anisotropic propagation, and tilted ice layers affect photon trajectories and arrival times~\citep{icecube2013transparency,icecube2025icemodel}.
Because IceCube's photosensors have nanosecond-scale timing resolution, simulations must reproduce not only the amount of detected light but also its complete arrival-time distribution.
Accurately modeling this response requires propagating large numbers of Cherenkov photons through a heterogeneous, strongly scattering and absorbing medium. 
Explicit Monte-Carlo (MC) propagation consequently makes photon tracking one of the dominant computational bottlenecks in IceCube's simulation pipeline~\citep{lundberg2007photonics,chirkin2019gpu}. 
A 2019 IceCube simulation campaign, for example, consumed approximately $100{,}000$ GPU-hours for photon-propagation simulations~\citep{sfiligoi2020preexascale}.

Modeling photon propagation in IceCube has historically followed two complementary MC-based strategies.
Photonics precomputes photon fluxes and arrival-time distributions and stores them in lookup tables~\citep{lundberg2007photonics}, whereas the Photon Propagation Code (PPC)---the direct, GPU-based MC simulator used as our reference throughout---tracks photons individually using a more detailed optical description~\citep{chirkin2013ppc}.
MC tables can also be represented by smooth, rapidly evaluated spline functions with analytic derivatives, as implemented in photospline~\citep{whitehorn2013photospline}.
However, this approach requires dense MC tables, and its coefficient grid grows rapidly with dimensionality: each medium model therefore requires a new table set and fit.

Learned surrogates and differentiable simulators provide complementary continuous alternatives.
\citet{lei2022implicit} use a SIREN-based neural field~\citep{sitzmann2020siren} to map a three-dimensional emission position to the time-integrated visibility of a fixed set of photosensors in the ICARUS~\citep{rubbia2011icarus} neutrino detector. 
Their model provides useful spatial gradients, but is tied to one sensor layout and does not represent source direction or photon arrival times. 
Another model, LUCiD, instead uses a differentiable ray tracer and demonstrates joint calibration and reconstruction~\citep{alterkait2026lucid}. 
Its current implementation targets homogeneous, analytically bounded detectors with surface-mounted sensors and combines expected charge with a first-arrival likelihood derived from a predicted timing distribution.
Neither approach addresses IceCube's full combination of directional sources, embedded receivers, and heterogeneous anisotropic transport.

We introduce \texttt{candela}: \textbf{C}omplex-media-\textbf{A}ware \textbf{N}etwork for \textbf{D}ifferentiable and \textbf{E}fficient \textbf{L}ight-transport \textbf{A}pproximation. \texttt{candela} is a SIREN-based implicit neural representation that learns the photon Green's function of the ice. 
Given the properties of a localized light source and the position of a receiver, the network predicts the expected detected light yield and its arrival-time distribution. 
This shared source--receiver kernel can be queried at arbitrary sensor positions, composed to simulate complete events, and differentiated end-to-end with respect to event parameters.
We demonstrate that \texttt{candela} accelerates IceCube photon propagation by $50$--$100\times$ relative to the standard MC simulation, with a cost that grows only weakly with neutrino energy. 
At the same time, it reproduces per-sensor photon arrival-time distributions to within the MC simulator's statistical uncertainty across $\sim$6 orders of magnitude in photon count.
Although designed and evaluated for IceCube, the same source--receiver formulation could be extended to other detectors and scientific instruments that require efficient, differentiable photon transport through complex scattering media.

\section{A neural field for photon transport}

\begin{figure}[t]
  \centering
  \includegraphics[width=0.94\linewidth]{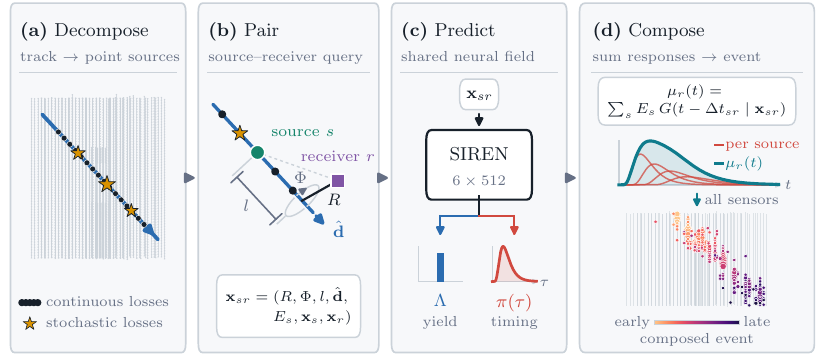}
  \caption{\textbf{The \texttt{candela} event-generation pipeline.}
  \textbf{(a)} Decompose a muon into point-like energy deposits.
  \textbf{(b)} Encode each source--receiver query by track-frame geometry $(R,\Phi,l)$, source direction and energy, and absolute positions.
  \textbf{(c)} \textbf{(c)} Predict per-GeV yield $\Lambda$ and timing density $\pi(\tau)$ for each query using one SIREN.
  \textbf{(d)} Scale each source response by its deposited energy and sum over sources to obtain the expected photon rate at every sensor, defining the complete event.}
  \label{fig:pipeline}
\end{figure}

The central observation behind \texttt{candela} is that photon transport is linear in the emitted light. 
A complete event can therefore be decomposed into localized energy deposits, the response of each deposit evaluated independently, and the resulting light fields superposed. 
This isolates transport through the heterogeneous ice from the event-specific pattern of particle trajectories and energy deposits.
Figure~\ref{fig:pipeline} summarizes the event-generation pipeline.

\subsection{The photon field of a point cascade}

The basic network input is the source--receiver query illustrated in Fig.~\ref{fig:pipeline}(b), denoted by $\mathbf{x}_{sr}$.
Consider a cascade source $s$ with position $\mathbf{x}_s$, direction $\hat{\mathbf d}$, and deposited energy $E_s$, together with a receiver at $\mathbf{x}_r$. 
For the displacement $\mathbf{r}_{sr}=\mathbf{x}_r-\mathbf{x}_s$, we describe the relative source--receiver geometry in a cylindrical frame aligned with $\hat{\mathbf d}$: $l$ is the displacement along the source direction, $R$ the perpendicular distance, and $\Phi$ the azimuth about the source axis. 
The network query additionally includes the absolute source and receiver positions because translation is not a symmetry of the detector: identical relative geometries can traverse different media due to the depth-dependence of optical absorption and scattering lengths in IceCube.

Arrival times are expressed using the residual coordinate $\tau=t-\Delta t_{sr}$, where $t$ is the absolute time and the anchor $\Delta t_{sr}=t_s+T_{sr}$ combines the emission time $t_s$ of the source with the unscattered geometric propagation time $T_{sr}$ to the receiver.
Removing this dominant time-of-flight dependence leaves the network to model delays and broadening induced by scattering. 
For each source--receiver query, the neural field represents an effective photon Green's function,
\begin{equation}
  G(\tau\mid\mathbf{x}_{sr})
    = \Lambda(\mathbf{x}_{sr})\,
      \pi(\tau\mid\mathbf{x}_{sr}),
  \label{eq:kernel-factorization}
\end{equation}
where $\Lambda$ is the expected detected photon yield per unit deposited energy and $\pi$ is a unit-normalized residual arrival-time density. 
We explicitly factor out the dominant linear energy scaling, so that a deposit of energy $E_s$ contributes $E_sG$. 
Source energy remains a conditioning variable to capture any residual nonlinearities in the photon yield.

\subsection{Network and training}

The shared neural field shown in Fig.~\ref{fig:pipeline}(c) maps every source--receiver query to two outputs: the integrated photon yield $\Lambda$ and arrival-time distribution $\pi$. 
Its backbone is a six-layer, width-512 SIREN, with separate heads for these two quantities.
We represent $\pi$ using an eight-component mixture of inverse-Gaussian distributions; its parameterization and time-support convention are described in Section~\ref{sec:timing}. 
The complete model contains approximately 1.6 million trainable parameters.

Training data are generated with PPC from isolated electromagnetic cascades spanning source position, direction, and energy. 
In total, we simulate $65{,}536$ source configurations. 
We repeat each configuration between once and approximately $100{,}000$ times to target a fixed total number of produced photons, with the number of repetitions determined by its energy.
The resulting sparse histograms use $64\times16\times96$ non-uniform spatial bins over $R\in[0,500]\,\mathrm{m}$, $\Phi\in[0,2\pi)$, and $l\in[-500,500]\,\mathrm{m}$, together with 256 residual-time bins over a $6\,\mu\mathrm{s}$ window.
The model is trained with a Poisson likelihood over these histograms, jointly constraining the integrated yield and arrival-time distribution.

\subsection{From point cascades to events}
\label{sec:compose}

A muon event is represented as a sequence of elementary cascade sources (Fig.~\ref{fig:pipeline}(a)). 
To generate its energy deposits, we use PROPOSAL~\citep{koehne2013proposal}.
As a muon traverses the ice, it loses energy continuously through many small interactions and stochastically through occasional large interactions that create localized cascades.
The continuous component is discretized along the track, while stochastic losses remain at their original positions.
Each source carries a deposited energy $E_s$, inherits the track direction, and is paired with every relevant receiver before the corresponding query $\mathbf{x}_{sr}$ is evaluated by the shared neural field. 
The expected photon rate at receiver $r$ is
\begin{equation}
  \mu_r(t)
    = \sum_s E_s\,
      G\!\left(t-\Delta t_{sr}\mid\mathbf{x}_{sr}\right),
  \label{eq:event-compose}
\end{equation}
where the anchor $\Delta t_{sr}$ is evaluated per source--receiver pair, so that each source's contribution enters at its own emission time and time of flight.
Expected photon counts are obtained by integrating $\mu_r(t)$ over the desired time bins.
All source--receiver pairs can be evaluated in parallel and summed to form the complete event (Fig.~\ref{fig:pipeline}(d)).

Feature construction, neural-field evaluation, time-bin integration, and event composition are implemented with differentiable tensor operations. 
For gradient-based applications, source responses are evaluated individually, preserving derivatives with respect to their continuous positions, directions, and deposited energies. 
The same model can therefore serve as both a fast event generator and a differentiable forward model for event reconstruction, a feature that we demonstrate in Section~\ref{sec:differentiability}.

\section{Results}

We evaluate \texttt{candela} on independently generated neutrino events spanning $10^2$--$10^6\,\mathrm{GeV}$. 
To isolate photon transport, each comparison fixes the particle-propagation and energy-loss realization and changes only the light-propagation method: \texttt{candela} or the GPU-based MC photon tracker PPC introduced above.
Repeated PPC propagations of these fixed losses additionally provide an MC-only reference for the irreducible error from finite photon statistics.

\paragraph{Event-level agreement}

\begin{figure}[t]
  \centering
  \includegraphics[width=0.94\linewidth]{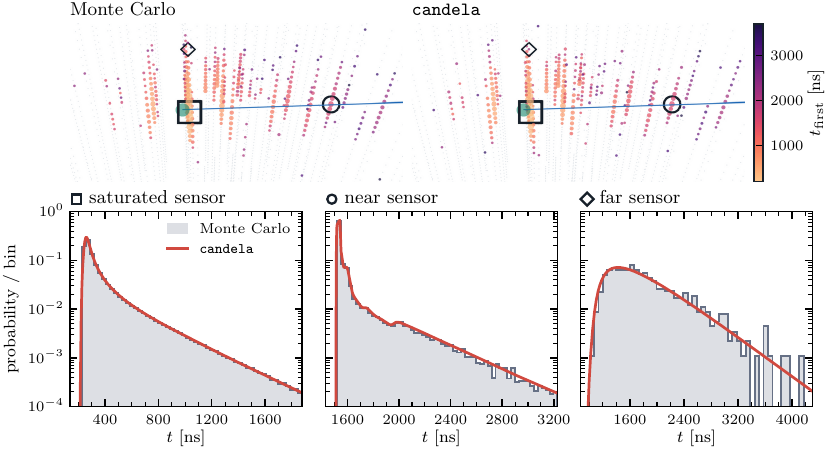}
  \caption{\textbf{Event-level agreement for a $100\,\mathrm{TeV}$ $\nu_\mu$ charged-current event.}
  \textbf{Top:} PPC (left) and \texttt{candela} (right) realizations of identical energy losses. The size and color encode photon count and first-hit time.
  \textbf{Bottom:} \texttt{candela} timing densities versus high-statistics PPC histograms at saturated, near-, and far-track sensors. Agreement spans both the prompt peak(s) and scattering tails.}
\label{fig:qualitative}
\end{figure}

Figure~\ref{fig:qualitative} compares the two simulators for a representative $100\,\mathrm{TeV}$ $\nu_\mu$ charged-current event. 
The surrogate reproduces both the spatial light deposition pattern and the progression of first-hit times across the detector. 
More quantitatively, the lower panels compare complete arrival-time distributions at the brightest sensor and at sensors near to and far from the track. 
Across these regimes, \texttt{candela} captures the prompt component, scattering-broadened structure, and long tail over four orders of magnitude in probability.

\paragraph{Speed and statistical fidelity}

\begin{figure}[t]
  \centering
  \includegraphics[width=0.94\linewidth]{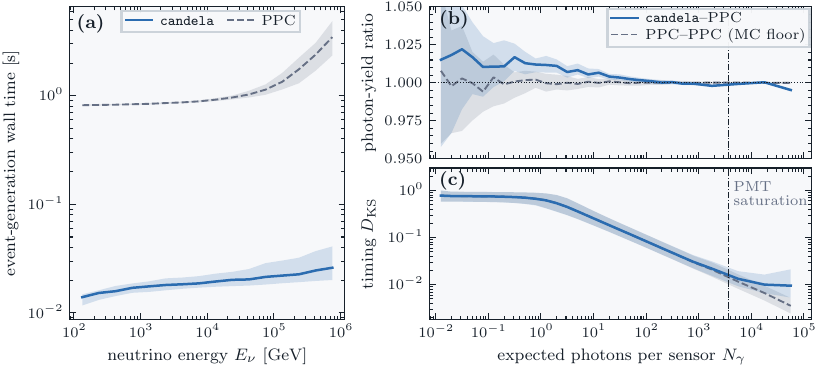}
  \caption{\textbf{Speed and fidelity on independent test events.}
  \textbf{(a)} Event-generation wall time for matched energy losses.
  \textbf{(b)} \texttt{candela} photon-yield prediction relative to the high-statistics PPC expectation.
  \textbf{(c)} KS distance to high-statistics PPC timing. 
  Lines and bands give bin medians and 16th--84th percentiles over events \textbf{(a)} or sensors \textbf{(b,c)}; PPC--PPC is the MC floor.}
\label{fig:results}
\end{figure}

Figure~\ref{fig:results}(a) compares event-generation wall time on the same NVIDIA RTX 5090 GPU.
PPC takes approximately one second at low energy and becomes increasingly expensive as the number of propagated photons grows. 
In contrast, \texttt{candela} requires only $10$--$20\,\mathrm{ms}$ and depends weakly on energy, yielding typical speedups of $50$--$100\times$. 
This scaling follows from replacing individual photons with batched evaluations of the learned source--receiver kernel, making the cost depend primarily on event geometry rather than photon multiplicity. 
We further exploit this compositional structure using a Lightcuts-inspired aggregation scheme~\citep{walter2005lightcuts}, described in Section~\ref{sec:lightcuts}.

To evaluate \texttt{candela} under realistic event-level photon fluctuations, we freeze the energy-deposition pattern produced by PROPOSAL for each event and propagate photons from the same depositions repeatedly with PPC. 
Pooling these re-simulations provides high-statistics estimates of the expected photon yield and arrival-time distribution at each sensor. 
We then generate two single-event realizations conditioned on the matched energy depositions: one with PPC and one sampled from the distributions predicted by \texttt{candela}. 
Each realization is compared with the high-statistics PPC reference. 
The single-PPC comparison therefore quantifies the irreducible variation from finite photon statistics, while the \texttt{candela} comparison additionally includes any surrogate error.
Figure~\ref{fig:results}(b) shows that the median \texttt{candela}-to-reference yield ratio remains within approximately $2\%$ of perfect agreement across the tested range of sensor brightness, closely tracking the PPC control. 
For timing, we compute the Kolmogorov--Smirnov (KS) distance $D_{\mathrm{KS}}$ between each single-event empirical arrival-time distribution and the high-statistics PPC distribution.
As shown in Fig.~\ref{fig:results}(c), \texttt{candela} follows the resulting PPC statistical floor over roughly 6 orders of magnitude in expected photon count. 
The vertical line marks the approximate onset of photomultiplier-tube (PMT) saturation, conservatively estimated from~\citep{Abbasi:2010vc}.
Above this intensity, the PMT response becomes nonlinear and unreliable.
Only in this brightest-sensor regime does \texttt{candela} separate visibly from the statistical floor.

\section{Conclusion}

We have introduced \texttt{candela}, a differentiable neural surrogate for photon transport through complex scattering media. 
By learning a shared source--receiver Green's function and linearly composing point-source responses, \texttt{candela} generates complete IceCube events $50$--$100\times$ faster than standard MC photon propagation while reproducing photon yields and arrival-time distributions to within the MC statistical limit in nearly all tested cases.
This combination of fidelity, computational efficiency, and differentiability provides a practical foundation for faster simulation and lays the groundwork for efficient gradient-based reconstruction and calibration in large-volume complex natural media. 
\paragraph{Limitations and outlook}
\texttt{candela} begins to separate from the MC statistical floor near the point of sensor saturation, a regime that is largely ignored in practice but may be relevant in future simulation-based studies.
The present study focuses on establishing the efficiency and fidelity of \texttt{candela}, with Section~\ref{sec:differentiability} providing a proof-of-concept for gradient-based event reconstruction.
Future work may investigate how to fully leverage the differentiability of \texttt{candela} for tasks such as event reconstruction and detector calibration.
Another promising extension is to additionally condition the neural field on optical parameters so that it can flexibly adapt to different media beyond IceCube.
With \texttt{candela}, we have proposed a type of neural surrogate that has the potential to help accelerate discovery of exciting scientific phenomena occurring in complex scattering media.


\section*{Acknowledgments}
This work is supported by the National Science Foundation under Cooperative Agreement PHY-2019786 (The NSF AI Institute for Artificial Intelligence and Fundamental Interactions, http://iaifi.org/). BTF is supported by the IAIFI Fellowship and Tayebati Fellowship. FJY is supported by the NSF Graduate Research Fellowship under Grant No. 2140743. CAA are supported by the Faculty of Arts and Sciences of Harvard University, the National Science Foundation (NSF, CAREER Grant No. 2239795), the John Templeton Foundation (Grant No. 63651), the Research Corporation for Science Advancement, and the David \& Lucile Packard Foundation. NK is supported by the National Science Foundation (NSF, CAREER Grant No. 2239795) and the David \& Lucile Packard Foundation.



\appendix

\section{Technical appendices and supplementary material}

\subsection{Arrival-time distributions}
\label{sec:timing}

Photon propagation through a scattering medium produces arrival-time distributions with a sharp onset followed by a long, geometry-dependent tail.
The inverse-Gaussian family provides a natural primitive for this behavior:
\begin{equation}
  f_{\mathrm{IG}}(u;\mu,\lambda)
  =
  \sqrt{\frac{\lambda}{2\pi u^3}}
  \exp\!\left[
      -\frac{\lambda(u-\mu)^2}{2\mu^2u}
  \right],
  \qquad u>0,
\end{equation}
where $\mu>0$ controls the characteristic arrival time and $\lambda>0$ controls the width. Its form, $u^{-3/2}\exp(-a/u-bu)$ up to a constant, is also characteristic of diffusive transport with absorption.

A single inverse Gaussian is insufficient to describe the range of prompt and multiply scattered propagation paths present in heterogeneous ice. 
We therefore represent the residual arrival-time density as an eight-component mixture,
\begin{equation}
  \pi(\tau\mid\mathbf{x}_{sr})
  =
  \sum_{j=1}^{8}
  p_j(\mathbf{x}_{sr})
  \frac{
      f_{\mathrm{IG}}\!\left(
          u(\tau);
          \mu_j(\mathbf{x}_{sr}),
          \lambda_j(\mathbf{x}_{sr})
      \right)
  }{
      Z_j(\mathbf{x}_{sr})
  },
\end{equation}
where $u(\tau)$ is a nonnegative shifted residual-time coordinate and $Z_j$ is the mass of component $j$ within the arrival-time range represented by the training data. 
The network predicts the component parameters and weights, with $p_j\geq0$ and $\sum_jp_j=1$. 
Since each component is normalized individually over the modeled range, $p_j$ directly represents the fraction of in-range photons associated with component $j$.

\subsection{Lightcuts}
\label{sec:lightcuts}

\begin{figure}[h]
  \centering
  \includegraphics[width=\linewidth]{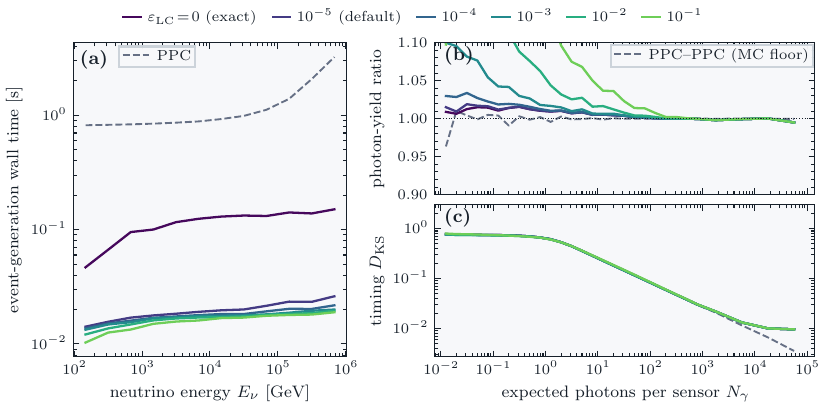}
  \caption{Performance of \texttt{candela} through a sweep of $\epsilon_{\mathrm{LC}}$ values.}
  \label{fig:lightcuts}
\end{figure}

A muon may contain thousands of elementary cascade sources, although many produce nearly indistinguishable responses at a given sensor.
Evaluating every source separately is unnecessary when their combined response can be approximated by a single representative query. 
Following the Lightcuts principle~\citep{walter2005lightcuts}, we order the sources along the track and organize them into a binary tree-like structure. 
The leaves are individual sources, adjacent groups are recursively merged, and the root represents the entire track. 
For each sensor, we traverse this hierarchy beginning at the root. 
A cluster is either retained as one query or divided into its two children, producing a sensor-specific adaptive cut.

A cluster \(C\) is represented by its energy-weighted centroid while preserving its total deposited energy. Its source-energy conditioning variable is the energy-weighted mean of the member energies rather than the cluster sum:
\begin{equation}
  E_C = \sum_{s \in C} E_s,
  \qquad
  \overline{\mathbf{x}}_C
  = \frac{1}{E_C}\sum_{s \in C} E_s \mathbf{x}_s,
  \qquad
  \overline{E}_C
  = \frac{1}{E_C}\sum_{s \in C} E_s^2 .
  \label{eq:lightcut-cluster}
\end{equation}
To determine whether a cluster should be divided, let \(\mu_C\) denote its predicted photon yield and let
\(\mu_C^{(2)}=\mu_{C_1}+\mu_{C_2}\) denote the sum of the predictions for its two children. 
A wide cluster can extend into a region of strong sensor response even when its centroid does not.
We therefore add a conservative probe \(\mu_C^{\mathrm{near}}\), evaluated at the point on the cluster's track interval closest to the sensor.
The cluster is divided when
\begin{equation}
  \max\left\{
      \left(\sqrt{\mu_C}-\sqrt{\mu_C^{(2)}}\right)^2,
      \left(\sqrt{\mu_C}-\sqrt{\mu_C^{\mathrm{near}}}\right)^2
  \right\}
  >
  \epsilon_{\mathrm{LC}} .
  \label{eq:lightcut-refinement}
\end{equation}
If the criterion is not met, the cluster is retained and its yield is corrected to the two-child estimate.
Because the threshold is expressed in expected detected photons, bright sensors are automatically refined more strongly than statistically insignificant ones. 
Setting \(\epsilon_{\mathrm{LC}}=0\) recovers exact source-by-source composition. 
As shown in Fig.~\ref{fig:lightcuts}, the default value \(\epsilon_{\mathrm{LC}}=10^{-5}\) substantially reduces generation time while remaining indistinguishable from exact composition in photon yield and arrival-time fidelity.

\subsection{Differentiability}
\label{sec:differentiability}

\begin{figure}[h]
  \centering
  \includegraphics[width=0.6\linewidth]{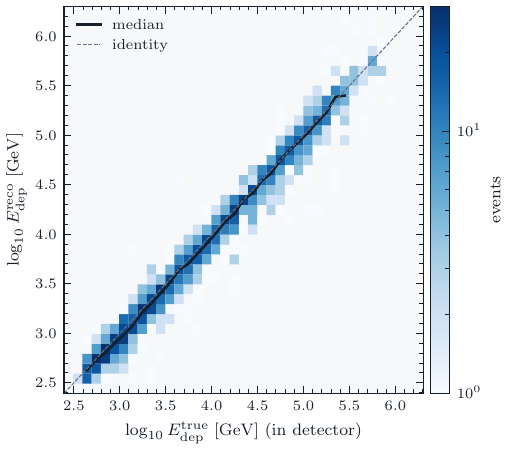}
  \caption{Demonstration of deposited energy reconstruction using the fully differentiable \texttt{candela}.}
  \label{fig:energy_reco}
\end{figure}

As a demonstration of gradient-based inference with \texttt{candela}, we reconstruct the energy deposited by through-going muons (Fig.~\ref{fig:energy_reco}). 
We simulate 1500 muons with energies drawn log-uniformly from $1\,\mathrm{TeV}$ to $1\,\mathrm{PeV}$ and isotropic directions, requiring each trajectory to cross the instrumented volume of an IceCube-like geometry. 
Energy losses are generated with PROPOSAL and propagated with PPC using the ice model employed to train \texttt{candela}; events with fewer than 20 detected photons are discarded. 
The target $E_{\mathrm{dep}}^{\mathrm{true}}$ is the sum of continuous and stochastic losses within the cylindrical hull of the sensor array.

The track direction, impact point, and reference time are first obtained from a geometric line fit. 
Holding this geometry fixed, we model the deposition profile as a minimum-ionizing component, $\langle dE/dx\rangle \simeq 0.26\,\mathrm{GeV\,m^{-1}}$, together with non-negative stochastic deposits $e_g$ in groups of adjacent track segments $g$.
Because \texttt{candela} predicts the photon yield per unit deposited energy,
the expected photon count at receiver $r$ is
\begin{equation}
\mu_r(\mathbf e)
= b_r + \sum_g A_{rg}(e_g)\,e_g ,
\end{equation} 
where $A_{rg}(e_g)$ is the predicted photon yield per GeV from segment group $g$ at receiver $r$, which retains a residual dependence on the deposit through the energy supplied to the network, and $b_r$ contains the fixed minimum-ionizing contributions. 
Given the observed count $n_r$, the deposits maximize the extended Poisson log-likelihood over all operational receivers, including those with $n_r=0$,
\begin{equation}
\mathcal{L}
= \sum_r
  \left[
    n_r\log\mu_r(\mathbf e)-\mu_r(\mathbf e)
  \right].
\end{equation}
The deposits are parameterized as $e_g = \exp\phi_g$, enforcing positivity, and the likelihood is maximized by gradient ascent on $\boldsymbol{\phi}$, with $\nabla_{\boldsymbol{\phi}}\mathcal{L}$ obtained by automatic differentiation end-to-end through the network.
Because the kernel yield per GeV depends weakly on the energy supplied to the network, $A_{rg}$ is recomputed using the fitted deposits and the optimization is repeated; two to three iterations are sufficient for convergence.
Figure~\ref{fig:energy_reco} compares the reconstructed and true in-volume deposited energies. 
The reconstruction closely follows the identity relation across the whole deposited energy range, with a median log-ratio of $-0.03$ and an interquartile range of $0.12$.


\begin{thebibliography}{99}

\bibitem[Alterkait et al.(2026)]{alterkait2026lucid}
Omar Alterkait, C\'{e}sar Jes\'{u}s-Valls, Ryo Matsumoto, Patrick de Perio, and
Kazuhiro Terao.
End-to-end differentiable calibration and reconstruction for optical particle
detectors. \emph{arXiv preprint
\href{https://arxiv.org/abs/2602.24129}{arXiv:2602.24129}}, 2026.

\bibitem[Chirkin(2013)]{chirkin2013ppc}
Dmitry Chirkin.
Photon tracking with GPUs in IceCube.
\emph{Nuclear Instruments and Methods in Physics Research Section A},
725:141--143, 2013.
doi:~\href{https://doi.org/10.1016/j.nima.2012.11.170}{10.1016/j.nima.2012.11.170}.

\bibitem[Chirkin et al.(2019)]{chirkin2019gpu}
Dmitry Chirkin, Juan Carlos D\'{i}az-V\'{e}lez, Claudio Kopper, Alexander R.~Olivas,
Benedikt Riedel, Martin Rongen, David Schultz, and Jakob van Santen.
Photon propagation using GPUs by the IceCube Neutrino Observatory.
In \emph{2019 15th International Conference on eScience (eScience)}, pages
388--393. IEEE, 2019. doi:~\href{https://doi.org/10.1109/eScience.2019.00050}{10.1109/eScience.2019.00050}.

\bibitem[IceCube Collaboration(2010)]{Abbasi:2010vc}
IceCube Collaboration.
Calibration and characterization of the IceCube photomultiplier tube.
\emph{Nuclear Instruments and Methods in Physics Research Section A},
618:139--152, 2010.
doi:~\href{https://doi.org/10.1016/j.nima.2010.03.102}{10.1016/j.nima.2010.03.102}.

\bibitem[IceCube Collaboration(2013)]{icecube2013transparency}
IceCube Collaboration.
Measurement of South Pole ice transparency with the IceCube LED calibration system.
\emph{Nuclear Instruments and Methods in Physics Research Section A},
711:73--89, 2013. doi:~\href{https://doi.org/10.1016/j.nima.2013.01.054}{10.1016/j.nima.2013.01.054}.

\bibitem[IceCube Collaboration(2017)]{icecube2017instrumentation}
IceCube Collaboration.
The IceCube Neutrino Observatory: instrumentation and online systems.
\emph{Journal of Instrumentation}, 12:P03012, 2017.
doi:~\href{https://doi.org/10.1088/1748-0221/12/03/P03012}{10.1088/1748-0221/12/03/P03012}.

\bibitem[IceCube Collaboration(2025)]{icecube2025icemodel}
IceCube Collaboration.
State of the ice model in the IceCube Observatory.
\emph{Proceedings of Science}, 501:1013, 2025.
doi:~\href{https://doi.org/10.22323/1.501.1013}{10.22323/1.501.1013}.

\bibitem[K\"ohne et al.(2013)]{koehne2013proposal}
Jan-Hendrik K\"ohne, Katharina Frantzen, Martin Schmitz, Tomasz Fuchs,
Wolfgang Rhode, Dmitry Chirkin, and Julia Becker Tjus.
PROPOSAL: A tool for propagation of charged leptons.
\emph{Computer Physics Communications}, 184(9):2070--2090, 2013.
doi:~\href{https://doi.org/10.1016/j.cpc.2013.04.001}
{10.1016/j.cpc.2013.04.001}.

\bibitem[Lei et al.(2022)]{lei2022implicit}
Minjie Lei, Ka Vang Tsang, Sean Gasiorowski, Chuan Li, Youssef Nashed,
Gianluca Petrillo, Olivia Piazza, Daniel Ratner, and Kazuhiro Terao.
Implicit neural representation as a differentiable surrogate for photon
propagation in a monolithic neutrino detector.
\emph{arXiv preprint
\href{https://arxiv.org/abs/2211.01505}{arXiv:2211.01505}}, 2022.

\bibitem[Lundberg et al.(2007)]{lundberg2007photonics}
Johan Lundberg, Predrag Miocinovic, Kurt Woschnagg, Thomas Burgess, Jenni Adams,
Stephan Hundertmark, Paolo Desiati, and Philipp Niessen.
Light tracking through ice and water---scattering and absorption in heterogeneous
media with PHOTONICS.
\emph{Nuclear Instruments and Methods in Physics Research Section A},
581(3):619--631, 2007. doi:~\href{https://doi.org/10.1016/j.nima.2007.07.143}{10.1016/j.nima.2007.07.143}.

\bibitem[Rubbia et al.(2011)]{rubbia2011icarus}
C.~Rubbia et al. (ICARUS Collaboration).
Underground operation of the ICARUS T600 LAr-TPC: first results.
\emph{Journal of Instrumentation}, 6(07):P07011, 2011.
doi:~\href{https://doi.org/10.1088/1748-0221/6/07/P07011}
{10.1088/1748-0221/6/07/P07011}.

\bibitem[Sfiligoi et al.(2020)]{sfiligoi2020preexascale}
Igor Sfiligoi, Frank W\"{u}rthwein, Benedikt Riedel, and David Schultz.
Running a pre-exascale, geographically distributed, multi-cloud scientific simulation.
In \emph{High Performance Computing}, volume 12151 of \emph{Lecture Notes in
Computer Science}, pages 23--40. Springer International Publishing, 2020.
doi:~\href{https://doi.org/10.1007/978-3-030-50743-5_2}{10.1007/978-3-030-50743-5\_2}.

\bibitem[Sitzmann et al.(2020)]{sitzmann2020siren}
Vincent Sitzmann, Julien N.~P. Martel, Alexander W.~Bergman, David B.~Lindell,
and Gordon Wetzstein.
Implicit neural representations with periodic activation functions.
In \emph{Advances in Neural Information Processing Systems}, volume 33,
pages 7462--7473, 2020.

\bibitem[Walter et al.(2005)]{walter2005lightcuts}
Bruce Walter, Sebastian Fernandez, Adam Arbree, Kavita Bala,
Michael Donikian, and Donald P.~Greenberg.
Lightcuts: A scalable approach to illumination.
\emph{ACM Transactions on Graphics}, 24(3):1098--1107, 2005.

\bibitem[Whitehorn et al.(2013)]{whitehorn2013photospline}
Nathan Whitehorn, Jakob van Santen, and Sven Lafebre.
Penalized splines for smooth representation of high-dimensional Monte Carlo
datasets. \emph{Computer Physics Communications}, 184(9):2214--2220, 2013.
doi:~\href{https://doi.org/10.1016/j.cpc.2013.04.008}{10.1016/j.cpc.2013.04.008}.

\end{thebibliography}
\end{document}